# Surface plasmon polariton waves with zero phase dispersion in a broad spectrum at Near-infrared wavelength

Shahram Moradi[1], Fazel Jahangiri[2]
1. Electrical and Computer Engineering, University of Victoria, Victoria, BC, Canada
2. Plasma and Laser Research Institute, Shahid Behesti University (SBU), Tehran, Iran
E-mail: photonicsandoptics@gmail.com

**Abstract**

We present theory to describe an engineering dispersion technique to obtain a broadband effective index near zero with an asymmetric planar photonic crystal. The theory provides the manipulating surface plasmon polariton (SPP) to provide alternating symmetric stacks of negative and positive effective indices. The odd alternating effective indices, including positive and negative refraction, arise from transverse resonance that depends on the geometry of the planar photonic crystal. The purposed technique remains wavepacket in zero phase dispersion since the created parity-time symmetries keep the phase constant in propagation direction. We use the plane wave expansion method to calculate band structure and transmission spectrum then validate with FDTD simulation. The results are compared to the recent experimental reports and they will be of significant interest to emerging applications in designing and fabricating metamaterials, optical filters, photonic sensors, photonic integrated circuits, near-field optics, and optofluidic biosensing applications.

**Introduction**

Engineering optical properties using artificial metamaterials composites provide plenty of intriguing features such as abnormal high-index[1], double-band negative index[2], zero refractive indexes[3] and so many others that come from the interaction between the radiation field with such metamaterials. In the recent study[4], the density of material undergoes perturbation effect and modifies refractive index and consequently the transmission spectra. Using this approach for the engineering material dispersion may be a new way to prevent wavepacket deformation[5]. The zero-refractive-index is a promising wave-manipulating platform[6–10] through which the propagating waves remain in a constant phase. This property is in high demand for light-based data communication where the wavepacket distortion is a critical concern. However, material dispersion regulates the wavepacket in a light path[11] but improper geometry may cause the phase distortion. Therefore, manipulating the geometry of dispersive media via perturbation effect is our approaches in this paper for engineering dispersion in the matter. However, the results in this article are beyond the limitations of today's fabrication but the recent study[4] provides a potential to breaks such fabrication barriers.

We get inspiration from the promising results in a recent study[4] for modifying the density of matter in a metamaterial (e.g. DNH) and consequently obtaining a smooth perturbation in the effective refractive index. We study continuous smooth perturbation in subwavelength of the incident that controls a broadband wavepacket from being distortion in a planar photonic crystal. Per the recently reported results[12] in the continuation of the previous study[13], the light flow undergoes $0^{th}$ order of phase accumulation in the propagation path due to a judicious alternating layers of gain and loss. This achievement introduces influences of perturbing a geometry as other potential applications of light-matter interaction[4] for establishing a broadband zero−$\hat{n}$ gap to remove some drawbacks in recent studies[12][13][14][15]. The dispersion with the zeroth-order mode may be a proper fit for preventing wavepacket distortion issues. It has been revealed that the dispersion engineering [16–19] in planar photonic crystals is controllable by means of tailoring geometry in order to adjust the phase, group and front velocities of the wavepacket[20]. Studying the effect of asymmetry in different topics [21–26] as a promising method of controlling light also provides a robust platform to conduct our design and simulation. In asymmetric lattices, wave packet undergoes a geometric variation in each cycle of iterative pairs (Berry phase) and forms an exotic dispersion relation[27,28] in specified directions. This occurs due to the manipulating the density of dielectric per volume[29] that accumulates phase gradually along the light propagation axis[12]. However, the gradient of dielectric in one dimension provides sections with an interface in between which requires a numerical method to solve Maxwell's equations[30]. The small interface between two regions with known solutions in each side of the region creates a delta function that may induce a phase transition abruptly[31,32]. As a result of this transition, the establishment of a $\mathcal{PT}-$ symmetry regarding the engineered exceptional points of degeneracy (EPD) in the interface of two perturbed pairs[11,33–36] is our suggested platform to initiate the design.

**Theory**

The discussion in this study regarding the zeroth order of stopband introduces an average zero effective refraction [37] using alternating stacks with odd values of effective refractive indices. This has been studied and demonstrated theoretically in [38] and experimentally in [13]. The zero stop bands unlike the Bragg grating[39] demonstrates $0^{th}$ order of phase accumulation. The experimental [13,40] results for the $0^{th}$ order of bandgap proves no shift in phase even if the length of the stacks gets variation through the light path. The promising result is developed in our previous study[12] via applying perturbing effect instead of using slab[13] in positive index stacks. In the recent study[12], we improved the attenuation of a signal based on the optical mismatch between the slabs and PhC layers that have been discussed in[12] comprehensively. The advantages of using all-dielectric [41] over metal-based composites[15] in our design is another reason to stop attenuating of signal through propagation. In addition, the narrow region of zero-index refraction that occurs only in a point ($\Gamma$) with Dirac cone band structure[14] [42] is the main concern of the presented paper. In this study, all previous issues including optical mismatch[13] in alternating layers, power attenuation[40], power absorption via metal-based composites[40] and more importantly the narrow region of zero refraction in [14] are obviated via continuous perturbing effect which we will discuss in the presented study. In other words, a chirality in distributing cylindrical holes by means of applying the perturbation effect utilizes to overcome some drawbacks in recent valuable reports[12][13][14][15].

Our suggested superlattice with the 1D binary periodic layers including NIM and PIM stack with a total optical path of $\Delta$ satisfies the Bragg gratings condition for the transfer matrix $(T)$ in which trace operator $(Tr)$ coalesces two eigenstates in a double layer $(Tr[T(\omega)] = 2\cos k\Delta)$ to form an EDPs for a particular range of frequencies[40]. By considering a perturbing effect in a linear custom, the refractive index may be a function of position $n(z)$ especially in a homogeneous medium. Thus, we can write the solution of the electric field for wave equation as a real function $E(z,t) = E_0 e^{i(\omega t - k_z)} \rightarrow E(z,t) = A(z) e^{i[\omega t - S(z)]}$ and $s(z)$ is a proper replacing function to the phase $k_z$ with respect to the consideration of the smooth variation of $n(z)$ with position to keep the coordination of the plane wave through the lattice. By plugging it into the wave equation in a homogeneous medium $\nabla^2 E = \frac{n^2}{c^2} \frac{\partial^2 E}{\partial t^2}$ and assuming WKB approximation, we can reach to the streamlined solution. As we assumed the slow variation of refractive index with the position to avoid any Fresnel reflection so:

$$s(z) = (\omega/c) \int_0^z n(z)dz = (\omega/c) . \Delta, \qquad (1)$$

In which $\Delta$ is the optical path in one cycle of propagation. Moreover, assuming $e^{ik(\omega)z}$ as a transfer function of the normalized plane wave transmitting in the **z**-axis, the parameters for the complex wave number such as $k(\omega) = \alpha(\omega) + j\beta(\omega)$ can be written in two either forms of $k' = \frac{\partial k(\omega)}{\partial \omega}$[12,13,40]:

$$\Delta_{NIM} = \int_1^0 n(\lambda, z)dz < 0 , k'(\omega, \vec{r}) < 0 \qquad (2)$$

, or

$$\Delta_{PIM} = \int_0^1 n(\lambda, z)dz > 0 , k'(\omega, \vec{r}) > 0 \qquad (3)$$

, in which $\Delta_{NIM}$ represents an optical path in one direction (NIM) and $\Delta_{PIM}$ refers to the same length but in the reverse direction (PIM). In addition, the average refractive index for double binary superlattice equals to $\bar{n}(z) = (1/\Delta) \int_0^\Delta n(z)dz$ in which the optical path for two components are identical ($\Delta = 2\Delta_{PIM} = 2\Delta_{NIM}$). As both components possess the same optical impedances ($\eta_{NIM} = \eta_{PIM}$) in all directions due to the equality of medium for both layers. Thus, the trace operator will be [40]:

$$Tr[T(\omega)] = 2\cos(\bar{n}\omega\Delta/c) - \left(\frac{\eta_{PIM}}{\eta_{NIM}} + \frac{\eta_{NIM}}{\eta_{PIM}} - 2\right) \times \sin\left(\frac{n_{PIM}\,\omega\,\Delta_{PIM}}{c}\right) \sin\left(\frac{n_{NIM}\,\omega\,\Delta_{NIM}}{c}\right) \qquad (4)$$

So, the solution for dispersion relation will be streamlined to: $Tr[T(\omega)] = 2\cos(s(z).\Delta)$, since the optical impedances possess the same value. Let us assume that $(\bar{n}(z).\omega.\Delta/c) = s(z).\Delta$ representing the proportionality of the phase component $k(\omega)$ to only the perturbing direction $s(z)$, and also consider a complex effective refractive index[12] $n_{eff}(z) = \omega \frac{d}{d\omega}\left(\left|\frac{c|\vec{k}(\omega,\vec{r})|}{\omega}\right|\right) + \left|\frac{c|\vec{k}(\omega,\vec{r})|}{\omega}\right|$ with $\mathcal{PT}-$ symmetry for the 1D binary double-layer superlattice.

As a result, the slope of band structure with respect to the gradient of $\omega$ will produce even values of $\alpha(\omega)$ and odd values for the imaginary part $\beta(\omega)$ in the complex wavenumber equation. Therefore, coalescence of two eigenvectors with same real values ensues under vanishing the imaginary parts, which means equality of phase transition for both layers.

Let us assume a propagated incident in the $\Gamma M$ direction of a hexagonal lattice with a lattice constant equal to $a = 0.5 \mu m$ and $r = 0.35\ \mu m$ as the radius of cylindrical holes undergoes variation of the radii($r/a$). Thus, it is possible to engineer the dispersion via perturbing unit cells $\Delta r = D_2 - D_1$ if we apply a smooth gradient ($\Delta r \ll a$) to avoid huge Fresnel reflection based on optical impedance mismatch. Due to the dependency of the wavevector to changing radii, the gradient of the wavevector with respect to the frequency ($\partial S(z)/\partial \omega$) will be either negative or positive in the dispersion diagram.

In Fig. 1, we compute the band structure for a regular hexagonal lattice (lattice constant: 0.5 [μm], radii: 0.35, silicon thickness: 0.32 [μm] and a substrate (silicon dioxide) with a thickness of 1 micrometer) and a perturbed lattice. We tune the volume of air through the lattice to achieve two layers with a different type of dispersion. The schematic of both ascendant and descendant in Fig. 1-**c**. shows the chirality in distributing the dielectric per volume. The ascendant in perturbing radii ($\Delta radii$: 0.35 → 0.37) that equals to 20 [$nm$] and the descendant in perturbing radii ($\Delta radii$: 0.35 → 0.33) with the same size 20 $nm$ produces two terminals. The name of terminals refers to the direction of the applied incident, which are either N or P component. A smooth linear growing of diameter to reach the radii equal to 0.37 forms P component (P: High to a low density of the dielectric distribution) and gradually reducing the diameter to reach radii equal to 0.33 forms N component (N: low to a high density of the dielectric distribution). Fig. 1-**a** demonstrates a computed band structure for the described geometrical parameters and its symmetric eigenvalues in the second Brillouin zone shown in Fig. 1-**b**. The aim of this section is to illustrate the effect of chiral distribution in printed cylindrical holes on the dispersion diagram.

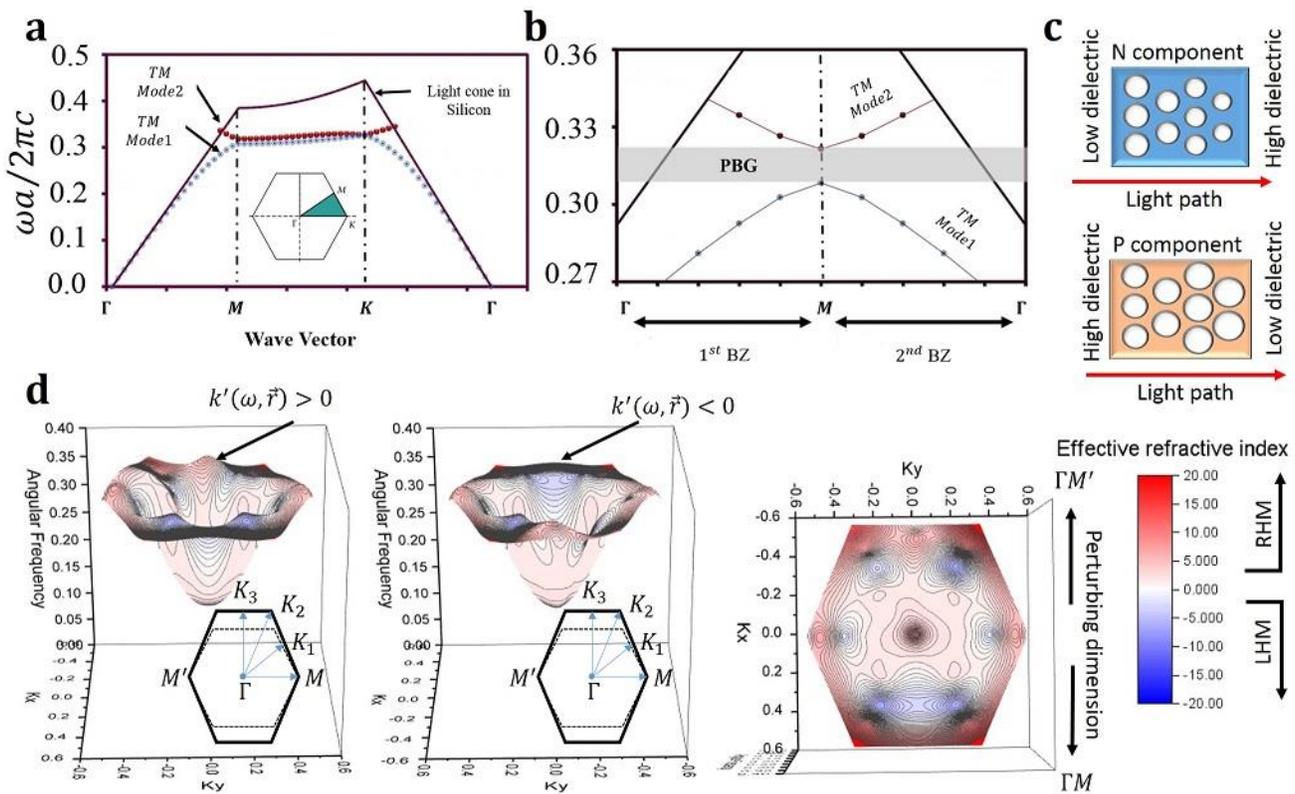

Fig. 1 | **Band structure and dispersion diagram in a regular and perturbed hexagonal lattice. a.** Band structure for a regular symmetric hexagonal lattice. **b.** Demonstrating symmetries in computed eigenvalues for the first and second Brillouin zone that illustrates no directionality in dispersion through the $\Gamma M$ direction. **c.** Schematic of chirality that perturbs lattice via the density of distributing dielectric per surface volume and consequently forming directionality in the material dispersion. **d.** Showing directionality in dispersion diagram in an asymmetric (perturbed) lattice via Riemann surface in which chromatic dispersion in two reverse directions gets either negative (LHM) or positive (RHM) effective refractive index due to odd gradient of the wavevector $k'(\omega, \vec{r})$ in one dimension.

We demonstrate opposite chromatic dispersion depending on the direction of perturbing path, which is shown in Fig. 1-**d**. In other words, establishing a two-terminal optical component with smooth contrast in one dimension of the lattice by means of compressing or expanding clusters[12] provides directionality in the light path. Furthermore, the suggested approach with different phase transitions based on the chosen terminal is promising enough to exploit it in a variety of applications such as optical modulation and switching purposes. However, we engage this component to form a balanced system with $0^{th}$ order of phase (zero-$\hat{n}$ gap[40]) for an active waveguide.

## Results and discussion

### 1. Zero-$\hat{n}$ gap

We construct 1D binary superlattice by means of a reverse combination of P and N components that provides a $0^{th}$ order of chromatic dispersion. Unlike the Bragg grating gap[37,38], the combination of two components with opposite refraction indexes [12][37][13]establishes a different type of stopband with net-zero phase accumulation at the end of the superlattice. Fig. 2 introduces a zero-$\hat{n}$ gap constructed of a balanced optical system via gain/loss parameters. First, we discuss some drawbacks in recently reported studies[10,12,40] to accentuate the priority of the suggested approach in this research. To begin with, heterogeneous stacks using in gain and loss parameter[13] with an optical mismatch $\eta_{NIM} \neq \eta_{PIM}$ causes tremendous attenuation of optical energy. Furthermore, a short spectrum frequencies in an accidental Dirac cone dispersion in the center of the Brillouin zone[10] makes it likely inapplicable. Moreover, the light absorption issue in reference[10] due to utilizing of the metal (gold) compromises its functionality. Albeit removing most of the mentioned issues in reference [12], the shorter bandwidth in zero-index regime still needs to be developed properly. To eliminate all aforementioned issues, using all-dielectric that satisfies these conditions in broadband is a critical step:

$|\Delta_{NIM}| = |\Delta_{PIM}|, l_{NIM} = l_{PIM}$ and $\eta_{NIM} = \eta_{PIM}$ (5)

However, the proposed criteria seem far-reaching, but the discussed two-terminal optical component satisfies all criterions in a broad range of frequencies.

We repeat the same parameters in reference [12] in Fig. 2 to examine the capability of swinging zero-$\hat{n}$ gap via modification of one parameter. The variation of radii from 0.35 in normalized transmission spectra in Fig. 2-**a** to 0.315 gives rise to swinging all zero-$\hat{n}$ gaps to the lower level of energy ($1/\lambda = 0.622$) that is shown in Fig. 2-**c**. In addition, the electric field distribution in front of each computed transmission spectrum admits $0^{th}$ order of dispersion after reaching the end of superlattice (Fig. 2-**b**, **d**). This is an initial inspiration result to broaden zero-$\hat{n}$ gap via perturbing one dimension ($\Gamma M$) that will be discussed in the rest of this study.

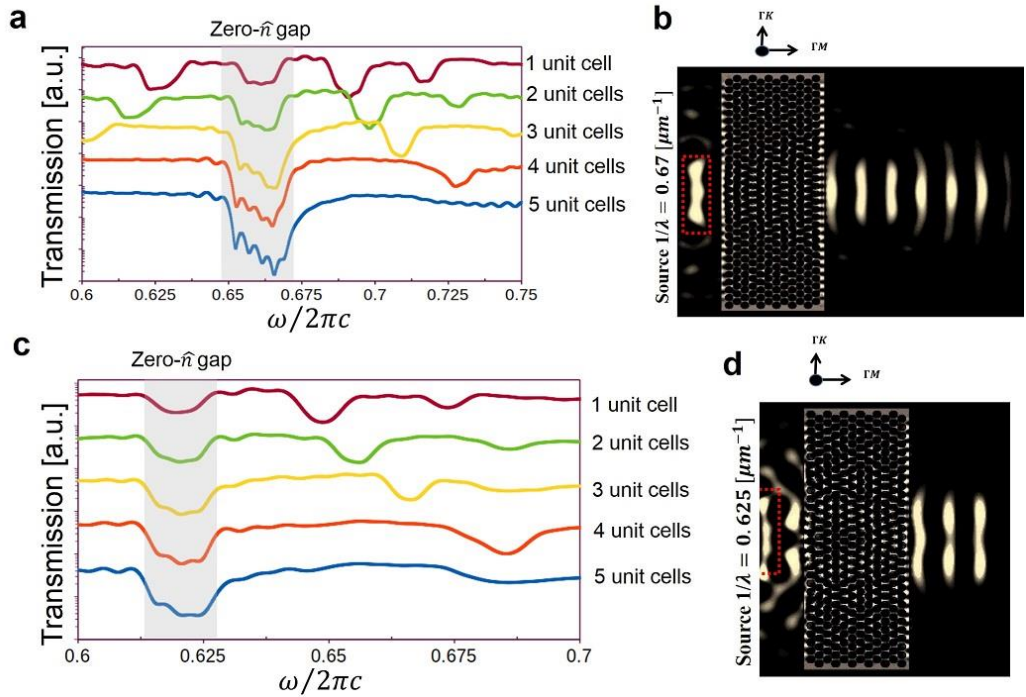

Fig. 2 | **Characterizing zero-$\hat{n}$ gap in a balanced Hamiltonian optical 1D binary superlattice. a.** Transmission spectrum for different unit cells with radii equals to 0.35 for both PIM and NIM layers and forming zero-$\hat{n}$ gap in $f_c = 0.665\ \mu m^{-1}$. **b.** Field intensity map (using linear grayscale) in a balanced superlattice in which applying a continuous source centering at $f_c = 0.67\ \mu m^{-1}$ in $\Gamma M$

direction for alternating stacks (1 unit cell) with one unit cell (radii:0.35). **c.** Transmission spectrum for different unit cells with radii equals to 0.315 for both PIM and NIM layers and forming zero-$\hat{n}$ gap at $f_c = 0.622\ \mu m^{-1}$. **d.** Field intensity map (using linear grayscale) with applying a continuous source centering at $f_c = 0.625\ \mu m^{-1}$ in $\Gamma M$ direction for alternating stacks (1 unit cell) with one unit cell (radii:0.315).

The net $0^{th}$ order of phase in the $\Gamma M$ direction gives rise to a harmonic chromatic dispersion through an established $\mathcal{PT}$-symmetric optical component whose balanced system is based on the symmetric arrangement of gain and loss parameters.

## 2. Broadening zero-$\hat{n}$ gap

The perturbed hexagonal array of cylindrical holes with $a = 0.5\ \mu m$ as a lattice constant printed on silicon $\varepsilon_r = 11.70$ with the thickness of $h_{silicon} = 0.32\ \mu m$ on insulator (silicon dioxide $\varepsilon_r = 2.07$ as a substrate with the thickness of $h_{SiO2} = 1\ \mu m$. The perturbing method in this study is an asymmetric distribution of dielectric via changing the depth, radius, and the lattice constant of cylindrical holes. This provides an equal length of stacks consisting $\{-\wedge/_2, 0\}$ and $\{0, +\wedge/_2\}$ through light path with alternating layers of negative and positive effective refractive index.

The chirality distribution requires breaking $\mathcal{PT}$-symmetry continuously in both layers including P or N component. However, the created layers possess similar optical properties except for the directionality of the effective refractive index. This is owing to the accumulation of shift in phase in every single cycle of the Berry phase transition through the continuous perturbed lattice. Here in the suggested non-Hermitian lattice, the accumulation of phase is linear and its magnitude in the time domain is proportional to the length and direction of the applied perturbation. We introduce a perturbing effect on the $\Gamma M$ direction to have a proportionality of phase on $\vec{r}$ with which the broken parity in lattice possesses time symmetry in the perturbed direction. The broader range of contrast in perturbation the broader formation of phase accumulation mismatch and consequently a broader band in the establishment of a zero-$\hat{n}$ gap. However, the problem with high perturbation size (comparable with lattice constant) is energy loss due to the Fresnel reflection in each geometrical transition. This approach creates a two-terminal optical component with which achieving both advance and delay phase is possible (Fig. 1-**a**). In addition, a comparable contrast of intensity in either output of the terminals for a particular range of frequencies is promising enough to exploit in plenty of switching and modulation applications (compare the outputs in Fig. 3-**b** and Fig. 3-**c**). The transmission spectrum in Fig. 3-**a** explains the reason for creation of two-terminal optical component based on either advance or delay shift in phase for an applied Gaussian pulse.

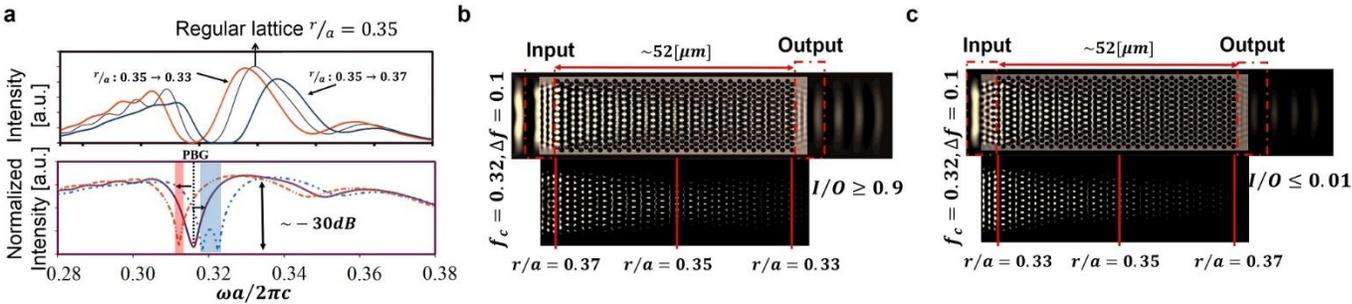

Fig. 3 | **Demonstrating directionality in the two-terminal optical component. a.** The up diagram shows a shift in phase for a perturbing lattice via only perturbing diameter of cylindrical holes and consequently the creation of either low to a high density of the dielectric ($^r/_a: 0.35 \rightarrow 0.33$) that causes a reddish shift in phase (shown with red color), and high to a low density of the dielectric ($^r/_a: 0.35 \rightarrow 0.37$) that causes a bluish shift in phase (shown with blue color). Furthermore, the normalized intensity (down) shows the depth of stopband that shifts along with the peak of a light pulse. **b.** Applying a light pulse with a given profile and its evolution through low to a high density of the dielectric; light flow through $\cong 52\mu m$ perturbed lattice (up) and showing field intensity map (using linear grayscale) along with the order of perturbing. **c.** Applying a light pulse with a given profile and its evolution through high to a low density of the dielectric; light flow through $\cong 52\mu m$ perturbed lattice (up) and showing field intensity map (using linear grayscale) along with the order of perturbing.

By considering a proper perturbation size (e.g. less than 100 nanometers in the whole length of a layer which is less than one-fifth of the lattice constant), the layers' characteristics including optical path, optical impedances, and the physical length get equal values. However, the band structure swings slightly up or down based on the type of perturbing effect. In -**a**, the broken $\mathcal{PT}$-symmetry forms two regions for a particular range of frequencies $\Delta \frac{\omega a}{2\pi c} \cong [0.305 - 0.335]$ with the opposite gradient of the wavevector. The numerical results for the effective complex refractive index $n^*_{eff} = n + j\kappa = |n|\angle\varphi$ in both P and N components demonstrate symmetric arguments of effective refractive index. The negative values of the effective refractive index in Fig. 4-b represents negative arguments and vice versa. This range of mismatching in effective refractive index depends on the engineering the density of the distributing dielectric per volume. However, the broadening the opposite band of refraction requires only 1D perturbation effect due to keeping the coordinate of plane wave constant. The results in Fig. 4-c demonstrate cancelation of the created opposite effective refractive index and consequently a broader zero-$\hat{n}$ gap. The Fig. 5 visualizes the light flow for a broadband Gaussian pulse in three-time intervals in which the light beam undergoes zeroth order of chromatic dispersion. The consecutive primary unit cells including P (low to high density in the distributing dielectric) and N (high to low density in the distributing dielectric) with a measured length of 14 [um] forms a binary 1D superlattice consists of seven alternating stacks.

The perturbing density of distributing dielectric in light path forms a broadband average zero-index in $\Gamma M$ direction with which the confining beam in the perpendicular direction conducts light flow with zigzagging rays through the 1D binary superlattice.

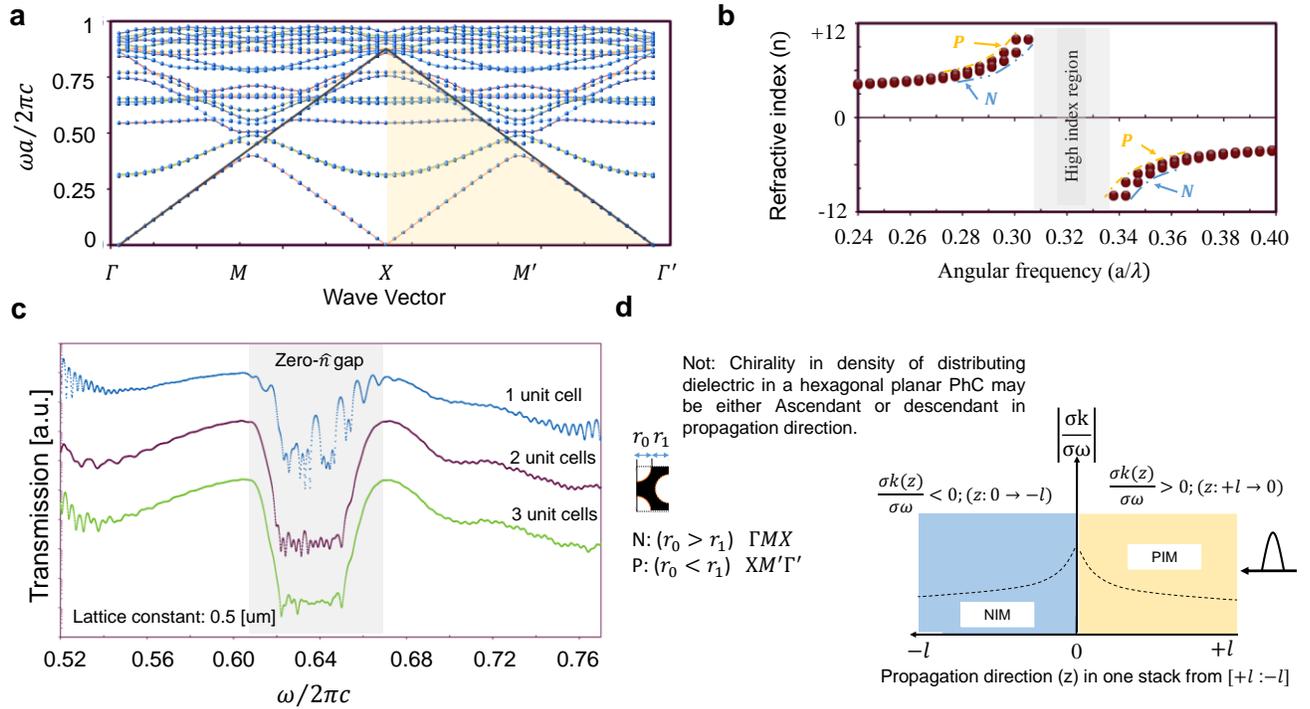

Fig. 4 | **Characteristics of a broad zero-$\hat{n}$ gap via perturbing effect. a.** Photonic band structure for superlattice with an inconstant gradient of wavevector with broken symmetries in radii. **b.** The first Brillouin zone and its pair symmetry in a perturbed hexagonal lattice. **c.** Effective refractive index for P (high to a low density of distributing dielectric) and N (low to a high density of distributing dielectric) which are the same lattice with different characteristics in phase (phase as time parameter in PT-Symmetry) due to odd values for $k'(\omega, \vec{r})$. **d.** The normalized transmission spectrum of a broad zero−$\hat{n}$ gap in 1.49~1.62 $\mu m$ wavelength interval that is shown in logarithm scale for three superlattices made of P-N component providing alternating layers of gain and loss. **e.** schematic of wavevector gradient (dashed lines) with respect to the addressed propagating direction from PIM (yellow) side starting from $+l$ to 0 and entering to the NIM (blue) layer till $-l$ point.

According to Zenneck[43], the mathematical solution of the wavevector in each cycle for 2D surfaces with alternating layers of $\varepsilon_{eff1} > 0$ and $\varepsilon_{eff2} < 0$ is $k = \frac{\omega}{c}\sqrt{\frac{\varepsilon_{eff1}\cdot\varepsilon_{eff2}}{\varepsilon_{eff1}+\varepsilon_{eff2}}}$. In other words, this allows an exponential bound mode to oscillate with elongation of wavevector if $|\varepsilon_{eff1}| \cong |\varepsilon_{eff2}|$ in the direction of radiation. Consequently, the wavelength reduces since the group velocity reaches near-zero values[44].

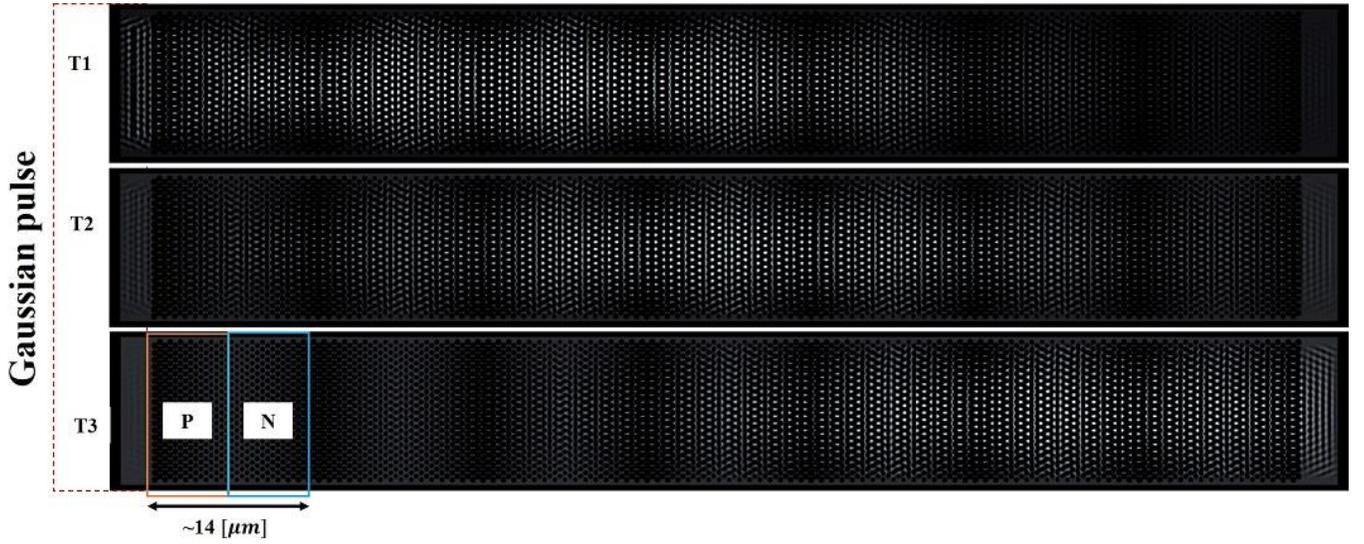

Fig. 5 | **Field intensity map (using linear grayscale) in 2D surface.** An amalgamation of two components P ($n_{eff1} > 0$) and N ($n_{eff1} < 0$) with different chromatic dispersions ($k'(\omega, \vec{r}) < 0$ or $k'(\omega, \vec{r}) > 0$) that forms $0^{th}$ order of effective refraction that is shown in three time intervals (T1, T2 and T3) in a lengthy ($\cong 100 [\mu m]$) P-N superlattice.

In Fig. 5, alternating stacks with an odd value of effective refractive index for a 2D slice of a 3D structure gives rise to elongation of wavevector and sharpening light pulse in only one direction. This achievement is promising to exploit in various application such as Distributed Bragg Reflectors (DBRs) in which the reflectivity enhances in a broadband region since the denominators depend on the configuration of layers. The laser beam that flows in a medium with $n_0$ experience ultra-high reflectivity with $R = \left[\frac{n_0(n_{eff2})^{2N} - n_s(n_{eff1})^{2N}}{n_0(n_{eff2})^{2N} + (n_{eff1})^{2N}}\right]^2$ that represents distributed two Bragg layers with different computed effective refractive index with odd values for two arranged layers $n_{eff1} = |n_{eff2}|$ on a substrate with $n_s$ [45,46].

## Conclusion

We presented a theory for broadening zero phase dispersion in a planar photonic crystal via breaking the symmetries in one dimension through a hexagonal planar photonic crystal. The described technique utilizes surface plasmon polariton waves to engineer dispersion to obtain both negative and positive effective indices with the same optical impedances. The importance of such achievement is broadening frequency band for gaining an average zero index dispersion to keep the phase constant through propagation direction. Comparison with both FED and FDTD simulations showed the promising predictive design capability of the theoretical approach and the results in this study were contrasted with both experimental and theoretical cases that have been published recently. The results in this article are relevant to zero phase dispersion, average zero index, and zero photonic bandgaps. However, in those works, the importance of broadening band for the zero index dispersion remains elusive. This work provides a better understanding of how to achieve a zero reflection coefficient in the near-zero effective index to boost the functionality of such a rare phenomenon at the experimental level. Knowledge of tailoring recent metamaterials to steer localized light is necessary for various light-based elements.